\documentclass[pre,preprint,superscriptaddress]{revtex4}
\usepackage{epsfig}
\usepackage{latexsym}
\begin {document}
\title {Oscillations and dynamics in a two-dimensional prey-predator system} 
\author{Ma\l gorzata Kowalik}
\affiliation{Department of Physics, A.~Mickiewicz University,
61-614 Pozna\'{n}, Poland}
\author{Adam Lipowski}
\affiliation{Department of Physics, A.~Mickiewicz University,
61-614 Pozna\'{n}, Poland}
\affiliation{Department of Physics, University of Geneva, CH 1211
Geneva 4, Switzerland}
\author{Antonio L.~Ferreira}
\affiliation{Departamento de Fisica, Universidade de Aveiro, 3810-193
 Aveiro, Portugal}
 %%%%%%%%%%%%%%%%%%%%%%%%%%%%%%%%%%%%%%%%%%%%%%%%%%%%%%%%%%%%%%%%%%%%%%%%%%%%%%%
\begin {abstract}
Using Monte Carlo simulations we study two-dimensional prey-predator systems.
Measuring the variance of densities of prey and predators on the triangular
lattice and on the lattice with eight neighbours, we conclude that temporal 
oscillations of these densities vanish in the thermodynamic limit.
This result suggests that such oscillations do not exist in 
two-dimensional models, at least when driven by local dynamics.
Depending on the control parameter, the model could be either in an active or
in an absorbing phase, which are separated by the critical point.
The critical behaviour of this model is studied using the dynamical 
Monte Carlo method.
This model has two dynamically nonsymmetric absorbing states.
In principle both absorbing states can be used for the 
analysis of the critical point.
However, dynamical simulations which start from the unstable absorbing state 
suffer from metastable-like effects, which sometimes renders the method 
inefficient.
\end{abstract}
\maketitle
%%%%%%%%%%%%%%%%%%%%%%%%%%%%%%%%%%%%%%%%%%%%%%%%%%%%%%%%%%%%%%%%%%%%%%%%%%%%%%%% 
\section{Introduction}
Nonequilibrium, many-body systems were usually regarded as one of the 
classical domains of physics.
However, recently is is becoming clear that such systems are of interest also 
in other disciplines such as biology~\cite{DROSSEL}, 
economy~\cite{STANLEY}, or sociology~\cite{STAUFFER}.
Consequently, notions of cooperative effects, formation of order or phase
transition, which were typically restricted to physics, proliferate other
scientific disciplines.

An example of multidisciplinary problem, which has application both in 
biological population dynamics and in hetero- or homogeneous catalysis, 
is the existence of temporarily periodic solutions in the so-called 
prey-predator systems.
A closely related problem is synchronization, which appears
in various biological contexts such as heart beats, wake-sleep cycles and
many other~\cite{GLASS}.
Earliest mathematical model of the temporal oscillations in interacting 
populations was proposed by Lotka and Volterra,
who described populations of preys and predators in terms of nonlinear 
differential equations~\cite{LOTKA}.
However, description of interacting populations in terms of ordinary 
differential equations, by necessity, introduces considerable simplifications.
In particular it neglects fluctuations and spatial inhomogeneities which
are certainly present in such systems.
One possibility to take such effects into account is to describe
such systems using lattice models.
The first approach of this kind was made by Chat\'e and Manneville~\cite{CHATE}.
Subsequently, more complicated versions of their model were 
studied~\cite{BOCCARA}.
Although in some cases these studies confirmed existence of periodic in 
time solutions, these models were driven by synchronous dynamics.
More realistic models with asynchronous dynamics were also 
examined~\cite{TANIA,PROVATA,TIBOR}.
As we already mentioned, closely related models are studied in the context of
catalysis~\cite{KORTLUKE}.

On general grounds one expects that fluctuations in low-dimensional systems 
are strong enough to destroy such temporal oscillations in the thermodynamic
limit.
An important question is: what is the critical dimension $d_c$ above which 
such oscillations will survive in the thermodynamic limit.
Certain arguments were given by Grinstein et al.~that $d_c=2$~\cite{GRINSTEIN}.
Numerical results for both synchronous~\cite{CHATE,HEMMINGSSON} and 
asynchronous models~\cite{LIP} seem to confirm that temporal 
oscillations exist but only for $d>2$.
Recently, however, numerical simulations by Rosenfeld et al. suggest that in 
a model with smart preys and predators
oscillatory behaviour might exist even in $d=2$ systems~\cite{ALBANO}.
They argued that the oscillatory phase is induced by a certain dynamical 
percolation transition.
If this percolative mechanism would be of more general validity, then the 
oscillatory phase should exist also in other two-dimensional models.
One would then argue that the fact that this phase was not observed in some 
earlier simulations~\cite{LIP} was
due to for example too small range of interactions of preys and predators.

In the present paper we examine two-dimensional versions of a prey-predator 
model~\cite{LIP} for the increased range of interaction.
However, our simulations show that even with the increased range of 
interaction, an amplitude of oscillations vanishes in the thermodynamic limit.
Let us notice that in three-dimensional version of the model, oscillations 
exists even for the nearest-neighbour interactions~\cite{LIP}.
Thus, our result strongly indicates that, at least for our model, there are
no oscillations in the case $d=2$.

Usually lattice prey-predator models have an absorbing state.
Consequently, they might undergo a phase transition in the steady state between
active and absorbing phases, which is expected to belong to the directed 
percolation universality class~\cite{DP}.
Our model is characterized by two absorbing states.
However, these states are asymmetric and only one of them is 
typically reached by the model's dynamics.
The second absorbing state can be considered as dynamically unstable.
Because of this asymmetry the model behaves as if it has only one
absorbing state and thus should exhibit directed percolation criticality, which
was confirmed for the one-dimensional case~\cite{LIPLIP}.
Critical behaviour of models with  absorbing states can be studied using the 
so-called dynamical Monte Carlo method~\cite{GRASS}.
In this method we prepare the system in an absorbing state and locally activate
it.
Statistics of such runs provides a very efficient method to study the critical
behaviour of such models.
This method was already applied to a number of models with a single 
absorbing state or with double but symmetric ones~\cite{HAYE}.
We are, however, not aware of studies for models with asymmetric 
absorbing states.
We show that when dynamical simulations start from an unstable absorbing 
state they suffer from metastable-like effects, which in the two-dimensional 
case renders the method inefficient.
In the one-dimensional case this effect is much weaker and both absorbing
states can be used to extract information about the critical point.

In section II we define the model, briefly mention the results of the simple
mean-filed approximations, and study the fluctuations of densities 
of populations.
In section III we present the results of the dynamical Monte Carlo method.
Section IV contains our conclusions.
%%%%%%%%%%%%%%%%%%%%%%%%%%%%%%%%%%%%%%%%%%%%%%%%%%%%%%%%%%%%%%%%
\section{Model and its steady-state simulations}
In our model on each site of a $d$-dimensional cartesian lattice of 
linear size $N$ we have a four-state variable $\epsilon_i=0,1,2,3$, which
corresponds to the site 
being empty ($\epsilon=0$), occupied by a prey ($\epsilon=1$), 
occupied by a predator ($\epsilon=2$) or occupied by a prey and a predator 
($\epsilon=3$).
Dynamics of this model is specified as follows~\cite{LIP}:\\
(i) Choose a site at random (say $i$-th).\\
(ii) With the probability $r$ ($0<r<1$) update a prey at the chosen 
site, provided that there is one (i.e., $\epsilon=1$ or 3); 
otherwise do nothing.
Provided that at least one neighbour of the chosen site is not 
occupied by a prey (ie., $\epsilon=0$ or 2), the prey (which is to be updated) 
produces one 
offspring and places it on the empty neighbouring site (if there are more 
empty sites, one of them is chosen randomly). 
Otherwise (i.e., when there is a prey on each neighbouring site) the prey 
does not breed (due to overcrowding).\\
(iii) With the probability $1-r$ update a predator at the chosen 
site, provided that there is one (i.e., $\epsilon=2$ or 3).
Provided that the chosen site is occupied by a predator but is not occupied 
by a prey ($\epsilon=2$), the predator dies (of hunger).
If there is a prey on that site (i.e., $\epsilon=3$), the predator 
survives and consumes the prey from the site it occupies.
If there is at least one neighbouring site which is not occupied by a 
predator, the predator produces one offspring and places it on the empty 
site (chosen randomly when there are more such sites).\\

To complete the description of this model we have to specify what are the 
neighbouring sites, i.e., sites where offsprings can be placed.
In previous studies of this model these were just nearest 
neighbours~\cite{LIP,LIPLIP}.
In the present paper of our main concern are $d=2$ models with further 
(but finite) range.

Steady-state description of our model is given in terms of densities of preys 
$x$ and predators $y$ defined as
\begin{equation}
x=\frac{1}{N^d}\sum_{i}(\delta_{\epsilon_i,1}+\delta_{\epsilon_i,3}),  \ \ \ 
y=\frac{1}{N^d}\sum_{i}(\delta_{\epsilon_i,2}+\delta_{\epsilon_i,3}),
\label{xy}
\end{equation}
where summation is over all $N^d$ sites $i$ and $\delta$ is Kronecker's 
$\delta$-function.
From the above rules it follows that the model has two absorbing states.
The first one  (PREY) is filled with preys ($x=1,\ y=0$) and the second 
one (EMPTY) is empty ($x=y=0$).
For large enough $r$, both populations coexist and the model is in the active
phase ($x>0,\ y>0$).
When the update rate of preys $r$ decreases, their number become to small to
support predators.
For sufficiently small $r$ predators die out and the system quickly reaches the 
absorbing state PREY.
In the simplest case, when the neighbouring sites are nearest neighbours, the 
phase transition between active and absorbing phase was observed at positive
$r$ for $d=1,2,3$~\cite{LIPLIP}.
For $d=1$ (linear chain) Monte Carlo simulations show that, as expected, 
the phase transition belongs to the directed percolation universality class.

To get additional insight into the behaviour of the model we can use a 
mean-field approximation.
In its simplest version one describes the model solely in terms of densities
$x$ and $y$ (which are strictly speaking time averages quantities defined in
Eq.~(\ref{xy})).
From the dynamics of the model and typical mean-field assumptions 
one can easily arrive at the following equations~\cite{LIP,LIPLIP}:
\begin{equation}
\frac{dx}{dt} = rx(1-x^{w})-(1-r)xy,
\label{mfaa}
\end{equation}
\begin{equation}
 \frac{dy}{dt} = (1-r)xy(1-y^{w})-(1-r)y(1-x).
\label{mfab}
\end{equation}
where $w$ is the number of neighbouring sites (see the dynamical 
rule (ii) and (iii)).
Predictions of such an approximation is, however, substantially different
from the behaviour of the model as observed in Monte Carlo 
simulations~\cite{LIP}.
In particular, approximation (\ref{mfaa})-(\ref{mfab}) fails to predict a 
phase transition
between active and absorbing phases at positive $r$ and for any dimension.
Moreover, there is no indication of the oscillatory phase as observed in the 
$d=3$ case.

An improved version of the mean-field approximation can be obtained introducing
a third variable $z$, which denotes a density of sites occupied by a prey and 
a predator ($\epsilon=3$).
Then, the mean-field equations are written as:
\begin{equation}
\frac{dx}{dt} = rx(1-x^{w})-(1-r)z,
\label{mfa1a}
\end{equation}
\begin{equation}
\frac{dy}{dt} = (1-r)z(1-y^{w})-(1-r)(y-z),
\label{mfa1b}
\end{equation}
\begin{equation}
\frac{dz}{dt} = \frac{rx(1-x^w)(y-z)}{1-x}-\frac{(1-r)z(1+z-x-y)(1-y^w)}{1-y}
-(1-r)zy^w.
\label{mfa1}
\end{equation}
In the approximation (\ref{mfaa})-(\ref{mfab}) the density of sites that are
occupied by a prey and a predator is 
simply given by the product $xy$.
In the approximation (\ref{mfa1a})-(\ref{mfa1}) this is an independent 
variable, whose time evolution follows from the dynamical rules of the model.
Numerical solution of (\ref{mfa1a})-(\ref{mfa1}) shows that this 
approximation is indeed more accurate~\cite{KOWALIK}.
In particular, it predicts that for $w\geq 4$ (which could be interpreted as
$d\geq 2$) there is a range of $r$ in the active phase where nonvanishing 
oscillations exist.
For $w=4$ (i.e., square lattice) this is still at odd with Monte Carlo 
simulations~\cite{LIP}.
Let us emphasize that both Monte Carlo simulations and mean-field calculations
are essentially independent on the initial configuration (provided, of course, 
we do not start from an absorbing state).

To examine the oscillatory behaviour, one can measure the standard deviation 
of the densities defined as $\sigma_{x}=\sqrt{<(x-<x>)^2>}$ and analogously for
the densities of predators $y$.
The symbol $<...>$, denotes time average in the steady state.
It was found that for $N\rightarrow\infty$ 
$\sigma=\sigma_x$ always decreases to zero for $d=1,2$~\cite{LIP}.
However, for $d=3$ there is a certain range of $r$ in the active phase and 
such that $\sigma$ remains finite for $N\rightarrow\infty$.
Such a behaviour indicates that finite-amplitude oscillations survive in the
thermodynamic limit.
%%%--------------------------------------
\begin{figure}
\centerline{\epsfxsize=8cm 
\epsfbox{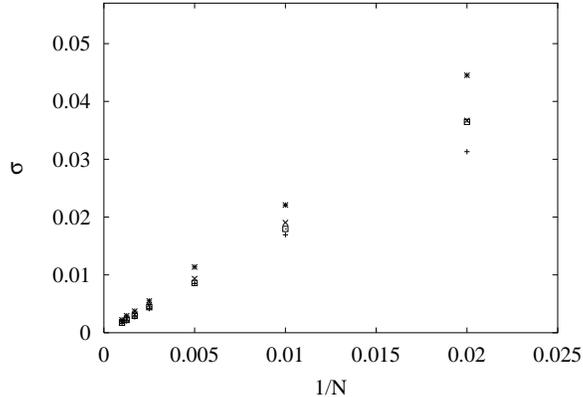}
}
%\figspace
\caption{The variance of the density of preys as a function of $1/N$ for 
the model on the triangular lattice.
Simulations were made for $r=0.4 (\Box), 0.3 (\star), 0.2 (\times),$ and 
0.15 (+).
}
\label{sigmatr}
\end{figure} 
%%%---------------------------------------

\begin{figure}
\centerline{\epsfxsize=8cm 
\epsfbox{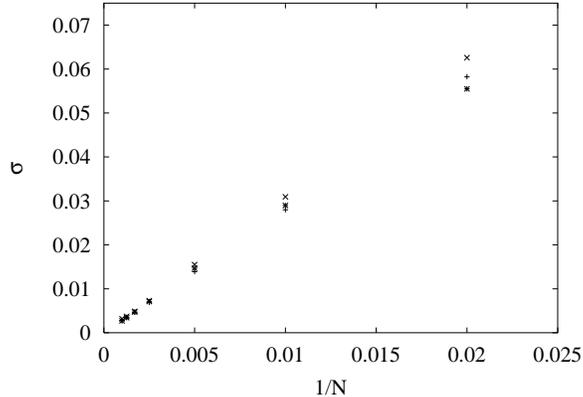}
}
%\figspace
\caption{The variance of the density of preys as a function of $1/N$ for 
the model on the lattice with nearest- and next-nearest neighbours.
Simulations were made for $r=0.35 (\star), 0.3 (\times),$ and 
0.25 (+).
}
\label{sigmaos}
\end{figure}

To check, whether an increased range of interactions might stabilize
oscillations in two-dimensional systems, we simulated our model on the 
triangular lattice and on the lattice with 8 neighbours (square lattice with 
nearest- and next-nearest neighbours).
On such lattices the parent site that is supposed to breed looks for an 
empty site among 6 or 8 neighbouring sites, respectively.
On such lattices we also observed a phase transition into absorbing phase 
around $r\sim 0.06-0.08$.
Similarly to other lattices~\cite{LIP}, for $r$ larger but close to the 
transition point fluctuations of densities exhibit oscillatory-like behaviour.
However, analysis of the variance of this fluctuations shows 
(see Fig.~\ref{sigmatr} and Fig.~\ref{sigmaos}) that in the
thermodynamic limit amplitude of oscillations vanishes.
Simulations were made for $N\leq 1000$ and simulation time was long enough 
to ensure that error bars are smaller than the plotted symbols.

Let us notice that for the lattice with 8 neighbours the coordination number 
is greater that in the three-dimensional (cubic lattice) case.
While in the former case oscillations die out in the thermodynamic limit, 
they survive in the latter one~\cite{LIP}.
It indicates that in two-dimensional systems oscillations are unlikely.
However, such a conclusion, if true, is applicable only to models equipped with
local dynamics.

One factor which is omitted in our approach is diffusion.
Our organisms diffuse  only effectively through the breeding process.
In related two-dimensional models, which were studied in the context of 
heterogeneous catalysis, diffusion was taken into account~\cite{KORT}.
It seems, however, that as long as the diffusion constant is finite, the
amplitude of oscillations vanishes in the thermodynamic limit.
%%%%----------------------------------------------
%%%%-----------------------------------------------
%%%%%%%%%%%%%%%%%%%%%%%%%%%%%%%%%%%%%%%%%%%%%%%%%%%%%%%%%%%%
\section{Dynamical Monte Carlo}
The essence of this method is to prepare the system in an absorbing state
and to activate it locally~\cite{GRASS}.
Quantities of the main interest are the probability $P(t)$ that the activity 
did not die out until time $t$ (the unit of time is defined as a single, 
on average, update of each site) and the average number of active sites $N(t)$
(averaged over all runs).
One expects, that at criticality these quantities have the power-law behaviour:
$P(t)\sim t^{-\delta},\ N(T)\sim t^{\eta}$, where $\delta$, and $\eta$ are
critical exponents characteristic to a given universality class.
Off  the critical point $P(t)$ and $N(t)$ deviate from the power-law behaviour.
%%%-----------------------------------------
\begin{figure}
\centerline{\epsfxsize=8cm 
\epsfbox{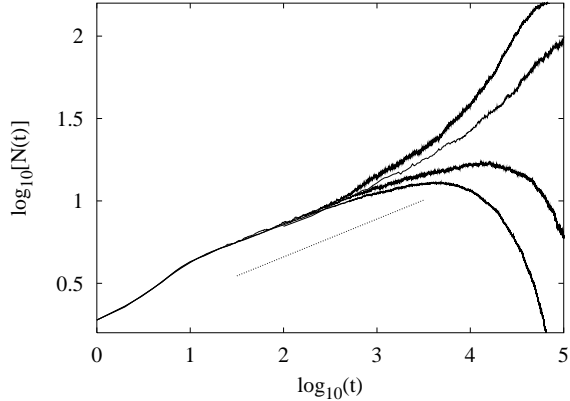}
}
%\figspace
\caption{The average number of predators $N(t)$ as a function of time $t$ for 
the model on the square lattice with the absorbing state PREY.
Simulations were made for (from top) $r=0.1099$, 0.1098, 0.1097 and 0.1096.
The dotted line has a slope 0.23 which corresponds to the exponent $\eta$ in
(2+1) DP~\cite{HAYE}.
}
\label{lognfinal}
\end{figure} 
%%%%%%--------------------------------------
\begin{figure}
\centerline{\epsfxsize=8cm 
\epsfbox{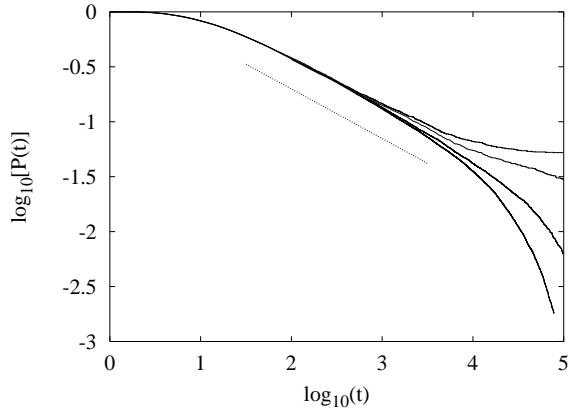}
}
%\figspace
\caption{The survival probability of predators $P(t)$ as a function 
of time $t$ for the model on the square lattice with the absorbing state PREY.
Simulations were made for (from top) $r=0.1099$, 0.1098, 0.1097 and 0.1096.
The dotted line has a slope 0.451 which corresponds to the exponent $\delta$ in
(2+1) DP~\cite{HAYE}.
}
\label{logpfinal}
\end{figure}
%%%%--------------------------------------------

We applied the dynamical Monte Carlo method to the two-dimensional model
with nearest neighbour interactions (square lattice).
In this case the steady-state calculations show that the model undergoes a 
transition at $r=r_c\sim 0.11$.
Our results in the  case of the absorbing state PREY are shown in 
Fig.~\ref{lognfinal}-Fig.~\ref{logpfinal}.
As an initial configuration we put all sites in a prey state ($\epsilon=1$), 
except a one, randomly selected, site which is in a prey-predator state
($\epsilon=3$).
Actually, in our model there are two quantities which scale approximately
in the same way, namely, the number of predators and the number of active sites 
(a site is active if it contains a predator, a prey and predator or a 
prey which is surrounded by at least one site without a prey).
To calculate $P(t)$ and $N(t)$ we considered only the number of predators (when
this number becomes zero we stop a given trial) but qualitatively 
the same results are obtained when the number of active sites is considered.
Using dynamical Monte Carlo method it is important to ensure that the 
spreading activity do not reach the boundary of the system.
We checked that the value which we used ($N=2000$) was sufficiently large.
From the behaviour of $P(t)$ and $N(t)$ we conclude that the phase
transition is located at $r=r_c=0.10975(5)$, which is far more accurate than 
the steady-state estimation.
Moreover, one can see that our data upon approaching the critical point
have a power-law behaviour with exponents consistent with (2+1) directed 
percolation~\cite{HAYE}.

\begin{figure}[!ht]
\centerline{\epsfxsize=8cm 
\epsfbox{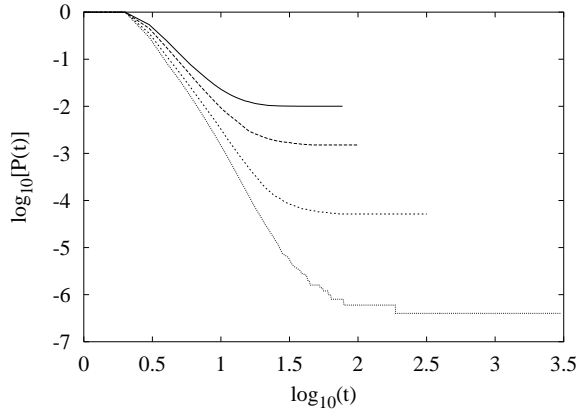}
}
%\figspace
\caption{The survival probability of predators $P(t)$ as a function 
of time $t$ for the model on the square lattice with the absorbing state 
EMPTY.
Simulations were made for (from top) $r=0.25$, 0.2, 0.15 and 0.12.
}
\label{ptpusty}
\end{figure}

\begin{figure}[!ht]
\centerline{\epsfxsize=8cm 
\epsfbox{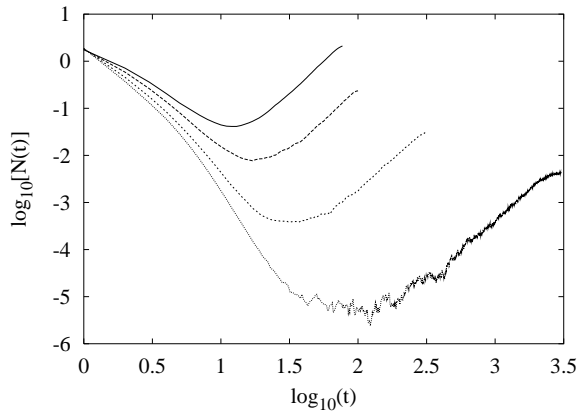}
}
%\figspace
\caption{The average number of predators $N(t)$ as a function 
of time $t$ for the model on the square lattice with the absorbing state 
EMPTY.
Simulations were made for (from top) $r=0.25$, 0.2, 0.15 and 0.12.
}
\label{ntpusty}
\end{figure}

Now, let us apply the dynamical Monte Carlo in the case of the EMPTY 
absorbing state.
As an initial configuration we put all sites in an empty state ($\epsilon=0$), 
except a one, randomly selected, site which is in a prey-predator state
($\epsilon=3$).
Numerical results are shown in Fig.~\ref{ptpusty}-Fig.~\ref{ntpusty}.
In these simulations the system is in the active phase ($r>r_c$) and relatively
far from the critical point.
Although the survival probability $P(t)$ saturates for large $t$ its asymptotic
values are very low (Fig.~\ref{ptpusty}).
For example, to confirm with this method that for $r=0.12$ the system is in 
the active phase tens of millions of runs must be made.
The reason for that is that for most of the runs predators die out very 
quickly.
Only very few runs generate sufficiently many preys which can support the 
population of predators.
If  the population of predators 
survives for a certain time, then, most likely, it will survive for an 
infinitely long time spreading through the system (see Fig.~\ref{ntpusty}).
We expect that such a behaviour exists for any $r>r_c$.
However, for $r$ close to $r_c$ the long-time survival events are extremely 
unlikely, which renders this method inefficient.
%%%%------------------------------------------
%%%%%----------------------------------------

Our model in this case resembles a metastable 
system~\cite{METAS}.
Such a system can exit the metastable state only through formation of a
sufficiently large seed of the stable phase.
In our case the EMPTY absorbing state corresponds for $r>r_c$ to a 
metastable state.
Only when a sufficiently large island of the active state occurs,
the system is irreversibly driven toward the active (stable) state.

\begin{figure}[!ht]
\centerline{\epsfxsize=8cm 
\epsfbox{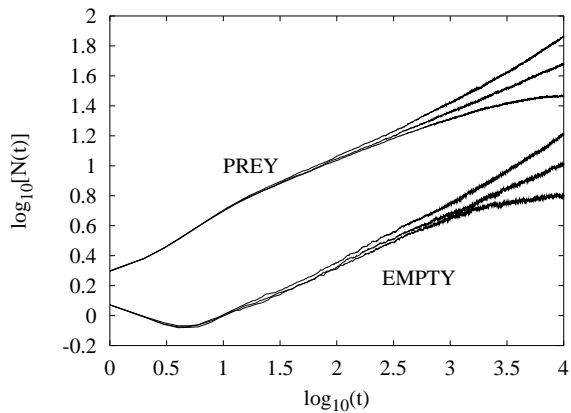}
}
%\figspace
\caption{The average number of predators $N(t)$ as a function 
of time $t$ for the model on the one-dimensional lattice with the absorbing 
state PREY and EMPTY.
For each absorbing state simulations were made for (from top) $r=0.493$, 0.491 
and 0.489.
The slope of the critical (central) lines is very close to the (1+1) directed 
percolation value $\eta=0.314$~\cite{HAYE}.
}
\label{dynd1}
\end{figure}

However, this metastable effect is much weaker in one-dimensional 
systems.
In this case the critical point is at $r=r_c=0.491(2)$~\cite{LIPLIP}.
Dynamical Monte Carlo simulations for both PREY and  EMPTY absorbing states
are relatively efficient (see Fig.~\ref{dynd1}).
In the case of the EMPTY absorbing state only a mild decrease of $N(t)$ is 
seen for small $t$.

We do not fully understand why metastability is much weaker in the $d=1$ case.
Most likely it is due to a much easier formation of a sufficiently large 
seed of a stable phase.
Outward spreading of such an active seed resembles a biased random walk and 
thus grows steadily in time.
In the $d=2$ case, the formation and growth of the active seed is more 
complicated.
%%%%%%---------------------------------------
%%%%--------------------------------------------
%%%%--------------------------------------------
%%%---------------------------------------------
%%%%%%%%%%%%%%%%%%%%%%%%%%%%%%%%%%%%%%%%%%%%%%%%%%%%%%%%%%%%%%%%%%%%%%%%
\section{Conclusions}
In the present paper we studied oscillatory and dynamical properties of the 
two-dimensional prey-predator systems.
Our results show that even for the increased range of interactions an
amplitude of oscillations vanishes in the thermodynamic limit.
It might suggests, as found by Grinstein et al. in a related 
context~\cite{GRINSTEIN}, that oscillations might persist but only for $d>2$.

In addition to that, we examined the validity and efficiency of the dynamical
Monte Carlo method for a model with two asymmetric absorbing states.
It turns out that for an unstable absorbing state in the two-dimensional
system this method is very inefficient.
On the other hand, both absorbing states can be used within this method in a
one-dimensional model.
%===============================================================================
\begin{acknowledgements}
A.~L.~F.~acknowledges financial support from the project POCTI/33141/99.
A.~L.~was partially supported by the Swiss National Science Foundation
and the project No.~OFES 00-0578 "COSYC OF SENS".
\end{acknowledgements}
%%%%%%%%%%%%%%%%%%%%%%%%%%%%%%%%%%%%%%%%%%%%%%%%%%%%%%%%%%%%%%%%%%%%%%%%%%%%%%%
%%%%%%%%%%%%%%%%%%%%%%%%%%%%%%%%%%%%%%%%%%%%%%%%%%%%%%%%%%%%%%%%%%%%%%%%%%%%%%%

%%%%%%%%%%%%%%%%%%%%%%%%%%%%%%%%%%%%%%%%%%%%%%%%%%%%%%%%%%%%%%%%%%%%%%%%%%%%%%%
%%%%%%%%%%%%%%%%%%%%%%%%%%%%%%%%%%%%%%%%%%%%%%%%%%%%%%%%%%%%%%%%%%%%%%%%%%%%%%%
\end {document}